\documentclass[]{jfm}

\usepackage{amsmath}     
\usepackage{amssymb}    
\usepackage{epsfig} 
\usepackage{natbib}

   \newcommand{\bk}{\mbox{\boldmath $k$}}

   \newcommand{\bu}{\mbox{\boldmath $u$}}

   \newcommand{\bx}{\mbox{\boldmath $x$}}

   \newcommand{\dee}[1]{\,{\rm d}{#1}}

   \newsavebox{\thalfbox}
   \sbox{\thalfbox}{$\textstyle\frac{1}{2}$}

   \newsavebox{\shalfbox}
   \sbox{\shalfbox}{$\scriptstyle\frac{1}{2}$}

\title[A model for rapid stochastic distortions of small-scale turbulence]{A model for rapid stochastic distortions of small-scale turbulence}

\author[B. Dubrulle, J.-P. Laval, S. Nazarenko and O. Zaboronski]{B.\ns  D\ls U\ls B\ls R\ls U\ls L\ls L\ls E$^1$,\ns J.-P.\ns L\ls A\ls V\ls A\ls L$^2$,\ns \\ S.\ns  N\ls A\ls Z\ls A\ls R\ls E\ls N\ls K\ls O$^3$\ns \and O.\ns  Z\ls A\ls B\ls O\ls R\ls O\ls N\ls S\ls K\ls I$^3$}

\affiliation{$^1$CNRS, URA 2464, GIT/SPEC/DRECAM/DSM, CEA Saclay 91191 Gif sur Yvette Cedex, France\\[\affilskip]
             $^2$CNRS, UMR 8107, Laboratoire de M\'ecanique de Lille, Blv. Paul Langevin, 59655 Villeneuve d'Ascq Cedex, France\\[\affilskip]
             $^3$Mathematics Institute, University of Warwick, Coventry CV4 7AL,United Kingdom}
\date{07 March 2003}

\begin{document}
\maketitle
\begin{abstract}
We present a model describing evolution of the small-scale
Navier-Stokes turbulence due to its stochastic distortions 
by much larger turbulent scales. This study is motivated by numerical
findings (\cite{laval01a}) that such interactions of separated scales play
important role in turbulence intermittency.
We introduce description of turbulence in terms of the moments
of the $k$-space quantities using a  method previously developed
for the kinematic dynamo problem (\cite{nazarenko03}).
Working with the $k$-space moments allows to introduce new useful
measures of intermittency such as the mean polarization and the
spectral flatness. Our study of the 2D turbulence shows that the
energy cascade is scale invariant and Gaussian whereas the enstrophy
cascade is intermittent. In 3D, we show that the statistics of 
turbulence wavepackets deviates from gaussianity toward dominance 
of the plane polarizations. Such turbulence is formed by ellipsoids
in the $k$-space centered at its origin and having one large, one 
neutral and one small axes with the velocity field pointing parallel
to the smallest axis.
\end{abstract}


\section{Introduction}
Finding a good turbulence model is a long standing problem.
To be useful in applications, the model has to be sufficiently simple 
and yet capable of capturing the basic physical processes such as the
energy cascade and the intermittent bursts. The cascades appear to be
a more robust property reasonably well described by classical turbulence
 closures such as the direct interaction approximation (DIA) (\cite{kraichnan61})
 and its
derivatives (e.g. EDQNM \cite{orszag66}).
 The turbulence intermittency appears to be a more 
subtle process which depends on the detailed features of the dynamical
fluid structures. Of particular importance 
is the question whether the intermittent bursts are caused by 
finite-time vorticity ``blow-ups'' (believed to become real singularities
in the limit of zero viscosity) or a ``slower'' exponential vortex
stretching by a large-scale strain collectively produced by the
surrounding vortex tubes. Note that the first process is local
in the scale space, - it is usually viewed as two or more vortex
tubes of similar and implosively decreasing radius. On the other
hand, the second process involves interaction of significantly
separated scales, - a thin vortex tube and a large-scale strain. 
Recent numerical simulations (\cite{laval01a}) 
indicate that it is the nonlocal scale interactions that are responsible for
the deviations of the structure functions from their Kolmogorov
self-similar value whereas the net effect of the local interactions is
to reduce these deviations. These conclusions lead to a model of turbulence
in which a  model closure (e.g. DIA) is used for the local interactions
whereas the nonlocal interactions are described by a wavepacket (WKB) 
formalism which exploits the scale separation. The later describes a
linear process of distortion of small-scale turbulence by a strain 
produced by large scales. Such a linear distortion is a familiar process
in engineering applications, for example, when a turbulent fluid flows through
a pipe with a sudden change in diameter. It is described by the rapid distortion
theory (RDT) introduced by \cite{batchelor54}. 
The model considered in this paper 
is different from the classical RDT in that the distorting strain is stochastic and,
therefore, we will call it the stochastic distortion theory (SDT). 
We will study a simplest  version of SDT in
which the large-scale strain is modeled by a Gaussian white (in time) 
noise in the spirit of the Kraichnan model used for turbulent passive
scalars (\cite{kraichnan74})  
and of the Kazantsev-Kraichnan model from the turbulent dynamo
theory (\cite{kazantsev68,kraichnan67b}).

 Because most of studies so far have focused on the theories with
local scale interactions, we will devote our attention mainly to the
description of the nonlocal interactions. The local interactions are
unimportant at small scales in 2D, but they should be taken into account
when 3D turbulence is considered. Similarly to RDT, the SDT model deals
with $k$-space quantities due to much greater simplicity of the pressure
term in the $k$-space. We will study statistical moments of the 
$k$-space quantities of all orders and not only the
second order correlators as it is customary for RDT.
The higher $k$-space correlators carry an information about the
turbulence statistics and intermittency which is not always
available from the two-point coordinate space correlators,
the structure functions, which are popular objects in the turbulence theory. 
Indeed, intermittency in some systems can be dominated by singular $k$-space 
structures which are not singular in the $x$-space (e.g. periodic fields).
These structures leave their signature on the scalings of the
 $k$-space moments but not on the $x$-space structure functions.
The system considered in the present paper is of this type, and another
example of this kind is the magnetic turbulence in the kinematic
dynamo problem (\cite{nazarenko03}). In fact, SDT bears a lot of similarities
to the turbulent dynamo problem and in this paper we will use a method
developed in \cite{nazarenko03} for its derivation. We will also see that, like in the
dynamo problem, the fourth order $k$-space moments allow us to introduce
the measures of the mean polarization and of the spectral flatness, - the quantities
of special importance for characterization of the small-scale turbulence.

\section{Stochastic distortion of turbulence}

%
Let us consider a velocity field in three-dimensional space that
consists of a component ${\bf U}$ with large characteristic
scale $L$ and a component $\bu$ with small characteristic 
scale $l$, $L \gg l$. In this case,  Navier-Stokes equation is
\begin{equation}
\partial_t {\bf U} + \partial_t \bu + ({\bf U} \cdot \nabla) {\bf U} +
({\bf U} \cdot \nabla) \bu + (\bu \cdot \nabla) {\bf U}
+ (\bu \cdot \nabla) \bu = - \nabla p +
\nu \nabla^2  {\bf U} + \nu \nabla^2  \bu. \label{ns}
\end{equation}
Let us define the Gabor transform (GT) (see \cite{nazarenko00b,nazarenko00c,nazarenko00d})
\begin{equation}
\hat{\bu} (\bx, \bk, t) = \int f(\epsilon^* |\bx  - \bx_0|)
e^{i \bk  \cdot (\bx  - \bx_0)} 
\bu (\bx_0, t) \dee{\bx_0},
\label{gt}
\end{equation}
where $1 \gg \epsilon^* \gg \epsilon$ and $f(x)$ is a function which
decreases rapidly at infinity, e.g. $\exp(-x^2)$.  Averaging
$\langle\cdot\rangle$ is performed over the statistics of a random
force which will be introduced below.

One can think of the GT as a local Fourier transform taken in a box
centered at $\bx $ and having a size which is intermediate between $L$
and $l$.  
The GT commutes with the time and space derivatives, $\partial_t$ and
$\nabla$.  Commutativity with $\partial_t$ is obvious.  Note that the
GT commutes with $\nabla$ only for distances from the boundaries which
are larger than the support of function $f$.
The inverse GT is simply an integration over all wavenumbers, e.g.
\begin{equation}
\bu (\bx, t) = {1 \over f(0)}
\int \hat{\bu} (\bx, \bk, t) {\dee{\bk} \over (2\pi)^3}.
\end{equation}
Here, we will study only the nonlocal interaction of small and
large scales and therefore we neglect the nonlinear term
$(\bu\cdot\nabla)\bu$ which corresponds to local interactions
among the small scales.
Let us apply the GT to the above equation with $k\sim
2\pi/l\sim 1 \gg 2 \pi /L \sim\epsilon$ and only retain terms up to
first power in $\epsilon$ and $\epsilon^*$ (we chose $\epsilon^*$ such
that $\epsilon^* \gg \epsilon \gg (\epsilon^*)^2$). All large-scale
terms (the first and the third ones on the LHS and the third one on
the RHS) give no contribution because their GT is
exponentially small. Equation for the GT of $\bu$ under such assumptions
where obtained in \cite{nazarenko00d}; it is

\begin{equation}
D_t \hat{\bu} + (\hat{\bu} \cdot \nabla) {\bf U} = { 2 \bk 
\over k^2} \hat{\bu} \cdot \nabla({\bf U} \cdot \bk )
 - \nu k^2 \hat{\bu}, \label{ss3}
\end{equation}
where
\begin{eqnarray}
D_t & = & \partial_t + \dot{\bx}\cdot\nabla + \dot{\bk}\cdot\nabla_{\bk},
\nonumber \\
\dot{\bx}  & = &  {\bf U}, \label{ray11}\\
\dot{\bk}  & = & - \nabla (\bk\cdot{\bf U}), 
\label{ray22} 
\end{eqnarray}
Equation (\ref{ss3}) provides an RDT description of turbulence
generalized to the case when both the mean strain  and the turbulence are
inhomogeneous.  This equation has the form of a WKB-type transport
equation with  characteristics given by (\ref{ray11}) and (\ref{ray22}).  
Consider this equation for a fluid path 
determined by $\dot {\bf x}(t) ={\bf U}$, so that $\hat {\bf u}({\bf k},{\bf x},t) \to 
\hat {\bf u}({\bf k},{\bf x}(t),t)$~\footnote{Hereafter, we drop hats on $\hat {\bf u}$
because only Gabor components will be considered. Also, we will not mention
explicitly dependence on the fluid path and simply write ${\bf u} \equiv {\bf u}
( {\bf k}, t)$. }
\begin{equation}
\partial_t u_m  = \sigma_{ij} k_i \, \partial_j
 u_m - \sigma_{mi} u_i + {2 \over k^2} k_m (\sigma_{ij} k_i u_j) - \nu k^2 u_m,
\label{beqn}
\end{equation}
where $\sigma_{ij} = \nabla_j U_i$ is the strain matrix and operators
$\nabla_i$ and $\partial_i$ mean derivatives with respect to $x_i$ and $k_i$
correspondingly ($i=1,...,D$). Note that strain $\sigma_{ij}$ (taken along a fluid path) 
enters this equation as a given function of time. 
 Equation (\ref{beqn}) is  applicable to  arbitrary
slowly varying in space large-scale flow. We  formulate the SDT model as equation
(\ref{beqn}) complemented by a prescribed statistics of the large-scale
flow. One can use, for example, a numerically computed large-scale
strain and use it as an input into the equation (\ref{ss3}) should
later be integrated numerically. In this paper, however, we would like to
derive a reduced model via the statistical averaging which is possible
by assuming a sufficiently simple statistics of the large-scale strain.
Experiments and numerical data indicate that Navier-Stokes turbulence
is Gaussian at large scales and we will use this property in our model.
We will further assume that the strain is white in time with \cite{nazarenko03} 
\begin{equation}
 \sigma_{ij} = \Omega (A_{ij} - {A_{ll} \over d} \delta_{ij})
\end{equation}
where $A_{ij}$ is a matrix the elements of which are statistically independent
and white in time,
\begin{equation}
 \langle A_{ij}(t) A_{kl}(0) \rangle = \delta_{ij} \delta_{kl}  \delta(t).
\label{Aij}
\end{equation}
This choice of strain ensure the incompressibility and statistical isotropy.
In this case
\begin{equation}
 \langle \sigma_{ij}(t) \sigma_{kl}(0) \rangle = \Omega (\delta_{ik} \delta_{jl} -
{1 \over d} \delta_{ij} \delta_{kl})  \delta(t).
\end{equation}
Note that this is not the only way to satisfy the incompressibility and the isotropy and 
there are  infinitely many ways to choose such statistics;
e.g. \cite{chertkov99}, \cite{falkovich01}, \cite{balkovsky99} choose
\begin{equation}
 \langle \sigma_{ij}(t) \sigma_{kl}(0) \rangle = \Omega ((d+1) \delta_{ik} \delta_{jl} -
\delta_{il} \delta_{jk} -\delta_{ij} \delta_{kl})  \delta(t).
\end{equation}
However, any such choice will lead to (up to a time-scale constant) the same final equations.
The Gaussian white in time  strain has also been used in the MHD dynamo theory
 (Kazantsev-Kraichnan model: \cite{kazantsev68}, \cite{kraichnan67b}) and in the theory of 
turbulent passive scalar 
(\cite{kraichnan74}; see also
 review of \cite{falkovich01}).  It  is a natural
starting point because of its simplicity.

Note that for realistic modeling of small-scale turbulence one has to describe a matching
to the large-scale range via a low-$k$  forcing or a boundary condition. As we will see
later, some properties
of the small-scale turbulence turn out to be independent of these effects, e.g. scalings
in 2D case, whereas the 3D case is more sensitive to the boundary conditions. 
Detailed modeling of the low-$k$ forcing/boundary conditions is beyond the scope of this paper.
Below, we will simply consider forcing-free evolution of a finite-support initial condition
(decaying turbulence). We will also consider finite-flux solutions in 2D corresponding to the 
turbulent cascades in forced turbulence.

%
\section{Generating Function}
%

Let us consider the following set of 1-point correlators of Gabor velocities,
\begin{equation}
\Psi^n_s =
 \langle |{\bf u}({\bf k})|^{(2n-4s)}  |{\bf u}({\bf k})^2|^{2s} \rangle               
\label{basis}
\end{equation}
with $n=1,2,3,...$ and $s=0,1,2,3,...$. Such correlators where shown in 
\cite{nazarenko03} to be a fundamental set in case of homogeneous isotropic
turbulence from which one can express any of two-point correlators of the following
kind
\begin{equation}
\langle u_{i_1}({\bf k}_1)  u_{i_2}({\bf k}_1) ...  u_{i_n}({\bf k}_1) 
u_{j_1}({\bf k}_2)  u_{j_2}({\bf k}_2) ...  u_{j_m}({\bf k}_2)  \rangle,
\end{equation}
where $i_1, i_2, .., i_n$ and $j_1, j_2, .., j_m$ take values 1, 2 or 3  indexing
the components in 3D space, and $n$ and $m$ are some arbitrary natural numbers.
Note that homogeneity and isotropy in SDT follow from the coordinate independence
and isotropy of the strain and it has to be understood only in a local sense, near
the considered fluid particle of the large-scale flow.

Now we define a generating function,
\begin{equation}
 Z(\lambda, \alpha, \beta, k) = \langle e^{\lambda|{\bf u({\bf k})}|^2 
+ \alpha {\bf u({\bf k})}^2 + \beta {\bf \overline u({\bf k})}^2}
 \rangle ,
\label{z} 
\end{equation}
where overline denotes the complex conjugation. This function allows one
to obtain any of the fundamental 1-point correlators (\ref{basis}) via differentiation
with respect to $\lambda, \alpha$ and $ \beta$,
\begin{equation}
\Psi^n_s =
 \left[ \partial_\lambda^{(2n-4s)}   \partial_\alpha ^{s}    \partial_\beta ^{s}  Z  
\right]_{\lambda= \alpha= \beta=0} .
\label{psi_v_z}
\end{equation}

To derive an evolution equation for $Z$ we will follow the technique developed in \cite{nazarenko03} for the turbulent dynamo problem.
Let us time differentiate the expression for $Z$ (\ref{z}) and use the
dynamical equation (\ref{beqn}); we have
\begin{eqnarray}
\dot Z = &  k_i \partial_j
 \langle \sigma_{ij} E \rangle -
\lambda \langle \sigma_{ml} ({\overline u_m} u_l + {\overline u_l} u_m) E
\rangle - 2 \alpha  \langle \sigma_{ml} { u_m} u_l E \rangle
- 2 \beta  \langle \sigma_{ml} {\overline u_m} {\overline u_l} E \rangle 
\nonumber \\
& - 2 \nu k^2 \langle ({\lambda|{\bf u({\bf k})}|^2 
+ \alpha {\bf u({\bf k})}^2 + \beta {\bf \overline u({\bf k})}^2}) E
\rangle
\label{zdot}
\end{eqnarray}
where 
\begin{equation}
 E = e^{\lambda|{\bf u({\bf k})}|^2 
+ \alpha {\bf u({\bf k})}^2 + \beta {\bf \overline u({\bf k})}^2} .
\label{E} 
\end{equation}
To find the correlators on the RHS of (\ref{zdot}), we use Gaussianity
of $\sigma_{ij}$
and perform a Gaussian integration by parts. Then, we use
whiteness of the strain field to find the response function
(functional derivative of $u_l$ with respect to $\sigma_{ij}$).
Finally, we use the isotropy of the strain so that the final equation
involves only $k = |{\bf k}|$ and no angular coordinates of the wave vector.
Leaving the derivation
for the Appendix, we write here only the final result,
\begin{eqnarray}
\dot Z &=& \Omega \big[ (1 - { 1 \over d}) k^2 Z_{kk} + 
{ 1 \over d} (4 {\cal D} + d^2 -1) k Z_{k} 
 +2(1-{2 \over d} +d) {\cal D} Z 
- {4 \over d} {\cal D}^2 Z  \nonumber \\ &&
 + 2 (\lambda^2 +4 \alpha \beta) Z_{\lambda \lambda} +
 2 \lambda^2 Z_{\alpha \beta}  + 8 \lambda \alpha Z_{\alpha \lambda}
+ 8 \lambda \beta Z_{\beta \lambda} +4 \alpha^2 Z_{\alpha \alpha} 
+4 \beta^2 Z_{\beta \beta}
\big] \nonumber \\ &&
-2 \nu k^2 {\cal D} Z,
\label{zdotF}
\end{eqnarray}
where the $k, \alpha, \beta$ and $\lambda$ subscripts in $Z$ denote 
differentiation with respect  $k, \alpha, \beta$ and $\lambda$
correspondingly and
\begin{equation}
{\cal D} = \lambda \partial_\lambda + \alpha \partial_\alpha +
\beta \partial_\beta.
\label{calD} 
\end{equation}
The number of independent variables in this equation
can be reduced by one taking into account that due to turbulence
homogeneity $Z$ depends on
$\alpha$ and $\beta$ only via combination $\eta = \alpha \beta$ (\cite{nazarenko03}).
We have
\begin{eqnarray}
\dot Z &=& \Omega \big[ (1 - { 1 \over d}) k^2 Z_{kk} + 
{ 1 \over d} (4 {\cal D} + d^2 -1) k Z_{k}  
 +2(1-{2 \over d} +d) {\cal D} Z 
- {4 \over d} {\cal D}^2 Z 
\nonumber \\
&& 
 + 2 (\lambda^2 +4 \eta) Z_{\lambda \lambda} +
 2 \lambda^2 (Z_{\eta} + \eta Z_{\eta \eta})  + 16 \lambda \eta Z_{\eta \lambda}
+ 8 \eta^2 Z_{\eta \eta} 
\big] 
-2 \nu k^2 {\cal D} Z,
\label{zdotFF}
\end{eqnarray}
where 
\begin{equation}
{\cal D} = \lambda \partial_\lambda + 2 \eta \partial_\eta.
\label{calD1} 
\end{equation}
Equation (\ref{zdotFF}) is the main equation of SDT. The RHS of this equation
describes interactions of the separated scales only. In
practical applications or numerical modeling one has to add to it a suitable
model for the local scale interactions. We leave this task for future and concentrate
below on studying the effect of the nonlocal interactions only.

%
\section{2D turbulence}
%

Let us first of all consider the 2D case. The large time dynamics of the small-scale
turbulence is known to be dominated in 2D by the nonlocal interactions due to
generation of intense large-scale vortices and, therefore, the SDT model
(\ref{zdotFF}) is relevant even without including a model for the local interactions.
 Note that equations for the 
2D turbulence in which only nonlocal interactions are left are formally
identical to the passive scalar equations. The energy spectra of the nonlocal 
2D turbulence and the passive scalars in the Batchelor regime were studied
in \cite{nazarenko00b} without making any assumptions on the strain statistics.
Here we will study the higher $k$-space correlators.

The 2D case is simpler than the 3D one in that all correlators 
$ \Psi^n_s $ with $s >0$ are not independent and can be expressed in terms
of $ \Psi^n_0 $ which we will call the energy series,
\begin{equation}
\Psi^n_0 \equiv E_n(k,t) =
\langle |{\bf u}({\bf k})|^{2n}  \rangle  =
 \left[ \partial_\lambda^{n}    Z  
\right]_{\lambda= \eta=0} .
\label{eser}
\end{equation}
The equations for correlators $E_n$ are to be obtained by differentiating
(\ref{zdotFF}) $n$ times with respect to $\lambda$ and taking it at 
$\eta =\lambda=0$ which gives
\begin{equation}
\dot E_n = {\Omega \over 2} \big[  k^2 (E_n)_{kk} + (3+4n) k (E_n)_{k} 
+ 4n (1+n) E_n \big]
-2 \nu n k^2 E_n.
\label{edot}
\end{equation}
 Of special interest are the cascade-type solutions realized when turbulence is
forced. To obtain these solutions we first re-write (\ref{edot}) as a continuity
equation in $k$-space; this can be done in two different
ways,
\begin{equation}
\partial_t (k^{2n-1} E_n)  = {\Omega \over 2} \big[  k^{-1} (k^{2n+2} E_n)_k  \big]_k
-2 \nu n k^{2n+1} E_n.
\label{ensdot}
\end{equation}
and
\begin{equation}
\partial_t (k^{2n+1} E_n)  = {\Omega \over 2} \big[  k^3(k^{2n} E_n)_k  \big]_k
-2 \nu n k^{2n+3} E_n.
\label{endot}
\end{equation}
In the absence of dissipation, $\nu =0$, equations  (\ref{endot}) and  (\ref{ensdot})
describe conservation of the quantities which have  spectral densities ${\cal E}_n = k^{2n-1} E_n$ and
${\cal F}_n = k^{2n+1} E_n$ respectively. For $n=1$ these quantities are just the energy and the
enstrophy~\footnote{Recall that we consider nonlocal turbulence and, although invariance
of enstrophy is easily seen from the vorticity conservation along the fluid paths,
the energy invariance is not obvious because there can be non-local energy exchanges
between turbulence and the large-scale flow. The conservation of energy was proved for 
initially {\em isotropic} turbulence in \cite{nazarenko00b}}. We will call invariants
${\cal E}_n$ and ${\cal F}_n$ the energy and the enstrophy series correspondingly.
The steady state solutions 
 in the range where viscosity is  negligible are 
\begin{equation}
E_n  = C_1^{(n)} \, k^{-2n}
\label{ec1}
\end{equation}
and 
\begin{equation}
E_n  = C_2^{(n)} \, k^{-2n-2},
\label{ec2}
\end{equation}
where $C_1^{(n)}$ and $C_2^{(n)}$  are arbitrary positive constants. For $n=1$ these solutions
where obtained in \cite{nazarenko00b}; solution (\ref{ec1}) corresponds to a constant energy flux 
  (and equipartition of the enstrophy) whereas
 (\ref{ec2}) corresponds to an enstrophy
cascade (and equipartition of the energy). Because the equations for  $E_n$ are
linear, any linear combination 
\begin{equation}
E_n  = C_1^{(n)} \, k^{-2n} + C_2^{(n)} \, k^{-2n-2}
\label{ec3}
\end{equation}
is also a stationary solution (in fact the general one). Note that the flux of invariant ${\cal E}_n$
is given by  $-{\Omega \over 2} k^{-1}(k^{2n+2} E_n)_k$ and it is always negative on solutions
(\ref{ec3}) whereas the flux of ${\cal F}_n$, which is $-{\Omega \over 2} k^3(k^{2n} E_n)_k$,
is always positive in the steady state. Thus, in forced turbulence solutions  (\ref{ec1}) will
form on the low-$k$ side of the forcing scale and solutions  (\ref{ec2}) on the high-$k$ side of it, 
which agrees with the general observation that the energy cascade is inverse and the enstrophy 
cascade is a direct one.

Let us introduce a spectral flatness,
\begin{equation}
F_n  = E_n / E_1^n.
\label{fl}
\end{equation}
For Gaussian fields, $F_n$ would be independent of $k$ and, therefore, $k$-dependence
of  $F_n$ bears an information about the scale invariance and presence of turbulence intermittency.
In particular, for the solutions  (\ref{ec1})  we have  $F_n = \hbox{const}$
indicating that the inverse cascades are not intermittent: turbulence produced
by a Gaussian forcing at some scale $k_f$ will remain Gaussian at $k<k_f$.
On solutions  (\ref{ec2}) we have $F_n \sim k ^{2n-2}$ which indicates broken
scale invariance and growing deviation from Gaussianity at small scales.
Such a small-scale intermittency in the direct cascades is due to nearly singular
$k$-structures having the shape of strongly elongated ellipses in 2D $k$-space which
are centered at the origin. Each strain realization will produce just one of these
(randomly oriented) ellipses out of an initial circular (isotropic) distribution.

%
\section{Nonlocal 3D turbulence}
%

3D Navier-Stokes turbulence is never likely to be nonlocal and for its
realistic description one should add a model of the local interactions
into the RHS of equation (\ref{zdotFF}). However, it is still interesting
to solve equation  (\ref{zdotFF}) as is in order to study the effect of pure
nonlocal interactions. We will see below that such a study
will reveal some interesting physics.

In the 3D case, equation  (\ref{zdotFF}) is
\begin{eqnarray}
\dot Z &=& {2 \Omega \over 3} \big[ k^2 Z_{kk} + 
4 k Z_{k}  + (8+2k \partial_k) {\cal D}  Z 
 + (\lambda^2 Z_{\lambda \lambda} +8 \eta^2 Z_{\eta \eta})  \nonumber \\ && 
 + (3 \lambda^2 - 4 \eta) ( Z_{\eta}  + \eta  Z_{\eta \eta}) 
+ 16 \lambda \eta Z_{\lambda \eta} +12 \eta Z_{\lambda \lambda} 
\big]
-2 \nu k^2 {\cal D} Z,
\label{zdotFFF}
\end{eqnarray}
The standard procedure to obtain equations for the correlators $ \Psi^n_s $
is to differentiate  (\ref{zdotFFF}) with respect to $\lambda$ and $\eta$ the 
required number of times and then to put $\lambda = \eta = 0$. In the 3D case,
all the correlators $ \Psi^n_s $ are independent and at each order $2n$ we
have a system of coupled equations rather than a single equation to solve
as it was the case in 2D. However, a decoupling arises asymptotically at
large times as we will see below. We will start by considering the second and
the fourth order correlators ($n=1$ and 2).

%
\subsection{Energy spectrum}
%

\begin{equation}
 E(k,t) \equiv  E_1(k,t) = 
\langle |{\bf u}({\bf k})|^{2}  \rangle  =
 \left[ \partial_\lambda    Z  
\right]_{\lambda= \eta =0} .
\label{enspec}
\end{equation}
Differentiating  (\ref{zdotFFF}) with respect to $\lambda$ 
and taking the result at $\lambda = \eta = 0$ we have
\begin{equation}
\dot E = {2 \Omega \over 3} (k^2 E_{kk} + 6 k E_{k} + 8 E) - 2 \nu k^2 E.
\label{Eeqn}
\end{equation}
This equation is similar to the
Kazantsev equation (\cite{kazantsev68}) describing evolution of the magnetic energy
spectrum in the kinematic dynamo theory.
Similarly to the dynamo theory, the total energy grows exponentially and therefore,
unlike the 2D case, no stationary cascade states are possible.
To have a steady state, one has to add a model of local interactions to SDT which will
be done in future publications. However, we will now study  equation (\ref{Eeqn})
to examine consequences of interactions of separated scales.

Recently, Schekochihin, Boldyrev and Kulsrud (\cite{schekochihin02a})
presented the  solution of 
the Kazantsev equation obtained by the Kontorovich-Lebedev transform and
we will use their results to solve (\ref{enspec}).
By substitution 
\begin{equation}
E  = e^{7  \Omega t/6} \, 
k^{-5/2} \, \phi(k/k_d , t) \;\;\; k_d = \sqrt{\Omega / 3 \nu },
\label{univ}
\end{equation}
one can reduce (\ref{Eeqn}) to
\begin{equation}
{3 \over 2 \Omega} \dot \phi(p,t) =  p^2 \phi_{pp} + 
 p \phi_p -p^2 \phi.
\label{phidot}
\end{equation}
The RHS of this equation is just the modified Bessel operator
and by using the Kontorovich-Lebedev transform one immediately gets 
for $t \gg 1 /\Omega$:
(\cite{schekochihin02a}, \cite{nazarenko03})
\begin{equation}
\phi(p,t) =  \hbox{const} 
\int_0^\infty ds \, s \, \hbox{sinh} (\pi s) \, K_{is} (p)  K_{is} (q) \,
e^{-s^2 2 \Omega t/3},
\label{phidis}
\end{equation}
where $K_{is}$ is the MacDonald function of an imaginary order
and  the constant  is fixed by 
 the initial condition.
At scales much greater than the dissipative one, $p \ll 1$, 
viscosity is not important  for $t \ll (\ln q)^2$ ($q \ll 1$ is the mean wavenumber
of the initial condition) and the solution (\ref{phidis}) becomes
(\cite{schekochihin02a,nazarenko03})
\begin{equation}
\phi  = \hbox{const} \, t^{-1/2} \, 
e^{ - {3 (\ln k/q)^2 \over 8 \Omega t}},
\label{phisoln}
\end{equation}
 This solution describes a spectrum with an expanding $k^{-5/2}$
scaling range (which means $k^{-1/2}$ for the one dimensional energy spectrum). 
At $t \sim (\ln q)^2$ the front of this scaling range reaches 
the dissipative scales and for $t \gg (\ln q)^2$ 
and ( \ref{phidis}) gives
\begin{equation}
\phi =  { \hbox{const} \over t^{3/2} } K_0(p).
\label{phiinf}
\end{equation}
Function $K_0(p)$ decays exponentially at large $p$ which corresponds
to a viscous cut-off of the spectrum.
For $p \ll 1$, $K_0(p) \approx - \ln p$, which means that
at large time the scales far larger than the dissipative one are
affected by viscosity via a logarithmic correction,
\begin{equation}
E(k) = { \hbox{const} \over t^{3/2} } e^{7  \Omega t/6}  k^{-5/2} \, \ln (k_d/k).
\label{logcor}
\end{equation}

\section{4th-order correlators, turbulence polarization and flatness}

There are two independent 4th order correlators 
\begin{equation}
S(k,t) = \Psi^{(2)}_0 =
 \langle |{\bf u}({\bf k})|^{4}   \rangle   
= \left[   Z_{\lambda \lambda} 
\right]_{\lambda = \eta=0}  \;\;\; \hbox{and}  \;\;\; 
T(k,t) = \Psi^{(2)}_1 =
 \langle |{\bf u}^2({\bf k})|^{2}   \rangle    
= \left[   Z_{\eta} 
\right]_{\lambda= \eta=0} 
\label{SandT}
\end{equation}
Differentiating (\ref{zdotFFF}) twice with respect to $\lambda$ and taking the result
at $\lambda= \eta=0$ we have
\begin{equation}
\dot S = {2 \Omega \over 3} (k^2 S_{kk} + 8k S_{k} +18 S + 6 T) - 4 \nu k^2 S.
\label{Seqn}
\end{equation}
Now, differentiating (\ref{zdotFF}) with respect to $\eta$ and taking the result
at $\lambda= \eta=0$ we get
\begin{equation}
\dot T = {2 \Omega \over 3} (k^2 T_{kk} +8k T_{k} + 12 T +12 S) - 4 \nu k^2 T.
\label{Teqn}
\end{equation}
By subtracting  (\ref{Teqn}) from  (\ref{Seqn}) we get a closed equation 
for $W=S-T$,
\begin{equation}
\dot W = {2 \Omega \over 3} (k^2 W_{kk}   +8k W_{k} +    6 W) - 4 \nu k^2 W.
\label{Weqn}
\end{equation}
The physical meaning of $W$ becomes clear if we re-write
it as \cite{nazarenko03}
\begin{equation}
 W = 4 \sum_{j \ne l}^3 \langle [\Im (u_j {\overline u_l} )]^2 \rangle =
 4 \sum_{j \ne l}^3 \langle |u_j|^2 |u_l|^2  \sin^2 (\phi_j-\phi_l) \rangle \ge 0,
\label{W}
\end{equation}
where symbol $\Im$ denotes the imaginary part and $\phi_j$ and $\phi_l$ are the phases 
of components $u_j$ and $u_l$ respectively. Thus, we see that $W$ contains information
not only about the amplitudes but also about the phases of the Fourier modes.
In particular,  $W \equiv 0$ corresponds to the case where all Fourier components of
the magnetic field have plane polarization. If $W \ne 0$ then other polarizations
(circular, ecliptic) are present. This is the case for example for Gaussian fields
when $W= E^2 /2 >0$. On the other hand, smallness of the phase differences can
be overpowered in  $W$ by large amplitudes. Therefore, a better measure of the mean
polarization would be a normalized $W$, e.g.,
\begin{equation}
P= W/S.
\label{P}
\end{equation}
Defined this way the mean turbulence polarization is an example of an important
physical quantity which can be obtained from the one-point Fourier correlators 
and is unavailable from the coordinate space (one-point or two-point) correlators.

Equation  (\ref{Weqn}) can be solved similarly to the energy spectrum equation  (\ref{Eeqn}),
namely by transforming it into equation  (\ref{phidot}) by a substitution similar to 
(\ref{univ}) and then using the solution  (\ref{phidis}).
In the inviscid regime ( $p \ll 1$, $t \ll (\ln q)^2$) we have
\begin{equation}
W = W_0 \, t^{-1/2} e^{-25 \Omega t /6} \, k^{-7/2} e^{ - {3 (\ln k/q)^2 \over 8 \Omega t}},
\label{Wsoln}
\end{equation}
where $W_0$ is a constant which can be found from the initial conditions.
We see that  $W$ develops a $ k^{-7/2}$ scaling range which is cut off at low and
high $k$ by exponentially propagating fronts.
Within this scaling range, $W$ decays exponentially in time.

Given $W$, one can find $S$ by using substitution $S=V+W/3$ which leads to 
a closed equation for $V$ which can be solved similarly to $E$ and $W$. This gives
for the inviscid regime 
\begin{equation}
S =  t^{-1/2}  \, k^{-7/2} e^{ - {3 (\ln k/q)^2 \over 8 \Omega t}}
\left( V_0 \, e^{47\Omega t/6 } + {1 \over 3} W_0 \,  e^{- 25 \Omega t /6} \right),
\label{Ssoln}
\end{equation}
where $V_0$ is another constant which can be found from the initial conditions.
For $t \gg 1$, the second term in the parenthesis should be neglected.
Then, we have the following solution for the mean turbulence polarization,
\begin{equation}
P =  W/S = {W_0 \over V_0} E^{- 12\Omega t}.
\label{Psoln}
\end{equation}
As we see, in the inviscid regime the mean polarization
tends to an independent of $k$ value which exponentially decays in time.
This means that all turbulence wavepackets  eventually become 
plane polarized. Recall that such turbulence is very far from being Gaussian
for which the mean polarization remains finite (elliptic and circular polarized
modes are present). 

In the diffusive regime, $t \gg (\ln q)^2$,  we have
\begin{equation}
W(k) =  W_0 \, t^{-1/2} e^{- 25 \Omega t /6} \, k^{-7/2} \, \ln (k_d/\sqrt{2} k)
\label{logcorW}
\end{equation}
and 
\begin{equation}
S(k) =  V_0 \, t^{-1/2} e^{47 \Omega t /6} \, k^{-7/2} \, \ln (k_d/\sqrt{2} k).
\label{logcorS}
\end{equation}
Thus, $P$ is still give by the same formula (\ref{Psoln})
indicating that the mean polarization
continues to further decrease in time  with the same exponential rate. 
Thus, by the time the diffusive regime is achieved $W$ can be essentially
put equal to zero.

The fact that the polarization becomes plane has quite simple physical explanation.
Indeed, a vorticity wavepacket of arbitrary polarization will be strongly
distorted by the stretching which mostly occurs along the dominant eigenvector of the
lagrangian deformation matrix (corresponding to the greatest Lyapunov exponent).
Such a stretching make any initial ``spiral'' flat for large time with the
dominant  field component lying in the plane passing through the stretching and the wavevector
directions (and, of course, ${\bf u} ({\bf k})$ is perpendicular to ${\bf k}$).

Another important measure of turbulence intermittency available from the 
$k$-space moments is the spectral flatness which
can be defined as $F=S/E^2$. For large time in the
inviscid regime
\begin{equation}
F \sim  t^{1/2} \, e^{-11 \Omega t/2} \, k^{3/2}.
\label{Fsoln}
\end{equation}
We see that the flatness growths both in time and in $k$ which
indicates presence of the small-scale intermittency. Such an intermittency
can be attributed to the presence of coherent structures in $k$-space.

One can also find solution for the fourth order correlator $S$ in the dissipative regime.
This will be done in the next section together with correlators of all higher orders.

\section{Large-time behavior of higher correlators}

The  observation in the end of the previous section (that there is a dominant
field component) allows us to predict that for large time
$|{\bf u}|^4 \approx |{\bf u}^2|^2 $ in each realization so that 
$ Z_{\lambda \lambda} \approx Z_{\alpha \beta}$. 
Therefore,  property $ Z_{\lambda \lambda} = Z_{\alpha \beta}$, 
if valid initially, should be
preserved by the equation for $Z$. Indeed, by using 
the equation (\ref{zdotFFF}) 
combination $w= Z_{\lambda \lambda} - Z_{\alpha \beta} = Z_{\lambda \lambda} -
Z_{\eta} - \eta Z_{\eta \eta} $ satisfies a closed homogeneous equation.
This means that  if $w \equiv 0$ at $t=0$ then it will remain identically
zero for any time. Thus, we can consider a class of solutions of
 (\ref{zdotFF}) (corresponding to large-time asymptotics of the general
solution) such that $ Z_{\lambda \lambda} = Z_{\alpha \beta}$.
Assuming this equality in   (\ref{zdotFF}) and putting $\eta=0$ we have,
\begin{equation}
\dot Z = {2 \Omega \over 3} \big[ k^2 Z_{kk} + 
4 k Z_{k}  + (8+2k \partial_k)  \lambda   Z_{ \lambda}  
 +4 \lambda^2 Z_{\lambda \lambda}  
\big]
-2\nu k^2  \lambda Z_{\lambda},
\label{zdotFFFF}
\end{equation}
This  gives the following  equations for the correlators
$E_n \equiv \langle |{\bf u}({\bf k})|^{2n}  \rangle$ is 
the correlator of order $2n$,
\begin{equation}
\dot E_n = {2 \Omega \over 3} \big[  k^2 (E_n)_{kk} + 
 (2 n + 4) k (E_n)_{k}  
 + 4n(n +1) E_n
\big]
-2 \nu k^2 n E_n.
\label{zdotDeg}
\end{equation}
Note that for $n=1$ this  coincides with 
for the energy spectrum equation (\ref{Eeqn}) which we have
already solved. Moreover, by substitution
\begin{equation}
E_n  = e^{(3n^2 +n -9/4) 2 \Omega t/3} \, 
k^{(-n-3/2)} \, \phi(k \sqrt{n} /k_d , t) 
\label{univ1}
\end{equation}
one can reduce (\ref{zdotDeg}) to an independent of $n$ equation (\ref{phidot}) 
for function $\phi$ the solution of which we already know.
For the inviscid 
regime ($1/\Omega \ll t \ll (\ln q)^2$) we have
\begin{equation}
E_n  = \hbox{const}(n) 
\, t^{-1/2} \, e^{(3n^2 +n -9/4) 2 \Omega t/3- {3 (\ln k/q)^2 \over 8 \Omega t}} \, 
k^{(-n-3/2)},
\label{Psi_n_pc}
\end{equation}
and in the diffusive regime ($t \gg (\ln q)^2$)
\begin{equation}
E_n  =  { \hbox{const}(n)  \over t^{3/2} } e^{(3n^2 +n -9/4) 2 \Omega t/3} \,k^{(n-3/2)} \,
 K_0( \sqrt{n} k /k_d).
\label{Psi_n_diff}
\end{equation}
This expression agrees with the results obtained in the previous sections
for $n=1$ and $n=2$.
We see that the main effect of the dissipation
is the prefactor change $ t^{-1/2} \to  t^{-3/2} $ 
 and the  $K_0(k/k_d)$ form-factor which corresponds
to a log-correction at $k \ll 1$ and an exponential cut-off at $k \sim k_d$.
Similar results were obtained for the magnetic field moments in \cite{nazarenko03}.
In that case it is the  exponential cut-off that causes the change of the exponential growth
in the mean magnetic energy (\cite{kazantsev68}, \cite{kulsrud92}) and in the higher $x$-space moments
of the magnetic field (\cite{chertkov99}).

The scalings in (\ref{Psi_n_pc}) and (\ref{Psi_n_diff}) with respect
to the order $n$ contain important information about the small-scale turbulence.
In particular the exponential growth in time with exponent 
 $\sim n^2$ indicates that the turbulence statistics is
log-normal. An equivalent result for the magnetic fields in the
kinematic dynamo problem was  obtained in \cite{chertkov99} and  \cite{schekochihin02b}.
The log-normality arises because the strain is a multiplicative noise for
the velocity field which becomes nearly one-dimensional
and because the time integrated strain tends to
become a Gaussian process. Formally, this result can be also obtained using the
random matrix theory of \cite{furstenberg63} (see also  \cite{falkovich01}), and a more
detailed physical explanation can be found in \cite{nazarenko03}.

The $k$-dependence of the Fourier correlators is also very important because it
gives an information about the dominant structures in the wavenumber space.
Suppose that initially the  turbulence is isotropic and concentrated in a ball
centered at the origin in the wavenumber space. For each realization such a ball will
stretch into an ellipsoid with one large, one short and one neutral dimensions.
One can visualize this ellipsoid as an elongated flat cactus leaf with thorns showing the
velocity field direction. Note that in this picture one component of the velocity field
(transverse to the cactus leaf) is dominant which is captured by the fact that the
polarization $W$ introduced in this paper tends to zero at large time.
Another consequence of this picture is that 
the wavenumber space will be covered by the ellipsoids more sparsely at large $k$ which implies
large intermittent fluctuations of the velocity field in the  $k$-space. 
These fluctuations could be quantified by the flatness $F$ which was shown 
in (\ref{Fsoln})  to grow as $ k^{3/2}$, a clear indication of the small-scale intermittency.

\section{Sensitivity of SDT to the strain statistics}

In the previous sections, the SDT model was formulated and studied assuming that
the large-scale strain is a Gaussian white noise process. This assumption, similar
to the Kazantsev-Kraichnan dynamo model, allowed us to obtain some important analytical 
solutions for the energy spectrum, the polarization, the flatness and higher-order
correlators which capture the turbulence intermittency. In real experimental and in numerical
turbulence the strain is generally far from being a Gaussian white noise. Thus, it is
interesting to study sensitivity of our SDT model and its predictions to the strain
statistics. 


In this section, we will numerically simulate  (\ref{beqn}) for two different types
of strain. First, we consider a synthetic Gaussian strain field with a finite correlation
time $\tau$ which is algorithmically generated as 
$$\sigma_{ij}(t+dt) = (1-dt/\tau) \, \sigma_{ij}(t) + \Omega  \sqrt{2\,  dt/ \tau} \left( A_{ij}(t) 
- {1 \over 3} A_{ll} \delta_{ij} \right)$$  where $A_{ij}$ is the same matrix as in (\ref{Aij}) and $dt$ is the 
time step.
The r.m.s. of the different strain components in this numerical experiment ranged from
2.9 to 3.6 and the correlation time was 0.02. Thus, the correlation time was about 16 times
less than the characteristic strain distortion time. We considered 512 strain realizations
of the synthetic field. 

In the second numerical experiment, the strain matrix components
were obtained from a $512^3$ spectral DNS of the Navier-Stokes equations at Reynolds number $R_\lambda \simeq 200$.
 The strain
time series were recorded along 512 fluid paths. The r.m.s. of the different strain components in this 
case ranged from 6.6 to 9.3 and the correlation time was approximately 0.08. Thus, the correlation
time in this case is of the same order as the inverse strain rate which is natural for the
real Navier-Stokes turbulence where both values are of order of the eddy turnover time at the
Kolmogorov scale.

Because of the fairly short correlation time and because of the Gaussianity, the first numerical 
experiment is closer to the Gaussian white-noise analytical model than the second experiment
where the strain is not Gaussian and has long time correlations. In a sense, comparison of results
of the first experiment and the analytical solutions
may be considered a performance test of the numerical method. On the other hand, comparison between the
results of the first and the second experiments allows us to establish their sensitivity to the
strain statistics. 

To compute (\ref{beqn}) we used a second order in time Runge-Kutta scheme with a time step of 
20 times less than the correlation time for the synthetic case. For the simulation with strain from DNS,
 a time step identical to the DNS have been used (i.e. 200 times less than the correlation time of strain)
In both numerical experiments, for each strain
realization we consider a distribution of wavepackets (2048 in the synthetic case and 8192 for the DNS)
with initial ${\bf k}$ chosen randomly in a sphere of radius $\vert{\bf k}\vert \simeq 2$. 
For each wavenumber ${\bf k}$, the two Fourier components of the velocity 
$u_1$ and $u_2$ are chosen randomly such that $u_j = \beta_j \gamma   e^{i 
2\pi\alpha_j}$ where $\alpha_j$, $\beta_j$ are uniform random numbers in the 
interval $[0,1]$ and  $\gamma=2.e-04$ is a constant (function of the total 
number of particles). The last component $u_3$ is  deduced from ${\bf k} 
\cdot {\bf u} = 0$ to respect the incompressibility. However, the minimum 
value of $\vert k_3 \vert$ is limited by the condition that when calculated 
$u_3$ appears to be excessively large such wavepacket is discarded.
The viscosity $\nu$ was set to $10^{-6}$ in both numerical experiments.


\begin{figure}[b]
\begin{center}
\includegraphics[width=.9\textwidth]{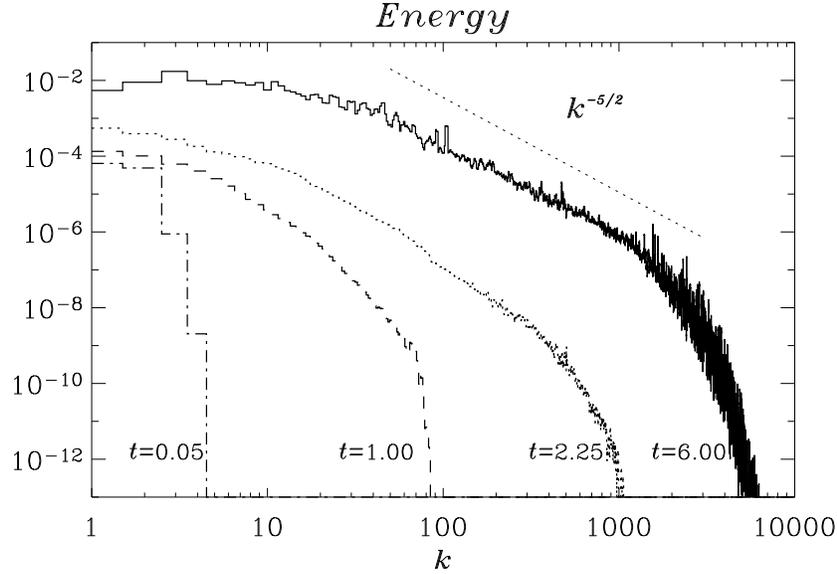}
\end{center}
\caption[]{Energy spectrum in the case of the synthetic Gaussian strain with a finite correlation time}
\label{fig1}
\end{figure}

\begin{figure}[b]
\begin{center}
\includegraphics[width=.9\textwidth]{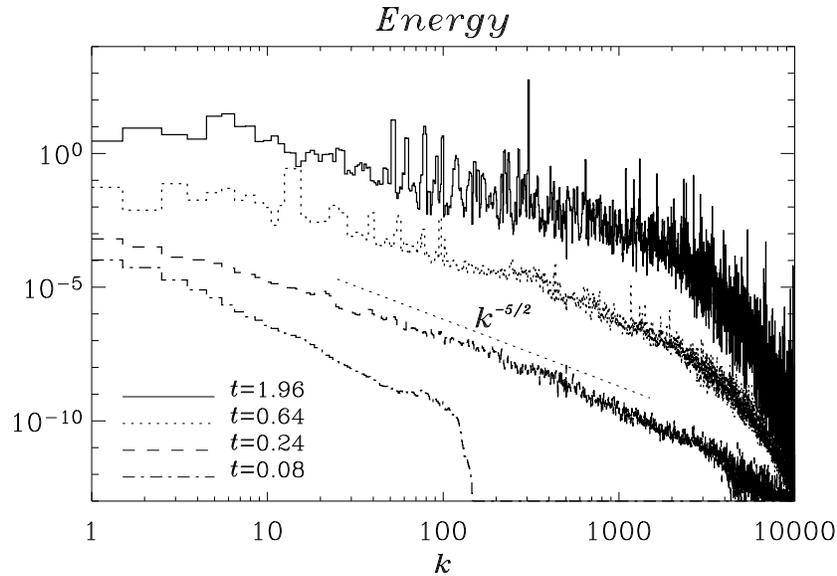}
\end{center}
\caption[]{Energy spectrum in the case of the strain obtained from 512$^3$ DNS}
\label{fig2}
\end{figure}

Figures \ref{fig1} and \ref{fig2} show the energy spectrum at several different moments of time
in the first and the second numerical experiment correspondingly. In both cases one can see an
excellent agreement with the theoretical -2.5 slope for time less than $t_d$ which is about
4.5 and 0.7 for the first and the second experiments respectively. At $t=t_d$, turbulence
reaches the dissipative scale and the spectrum accepts a log-corrected shape.

\begin{figure}
\begin{center}
\includegraphics[width=.9\textwidth]{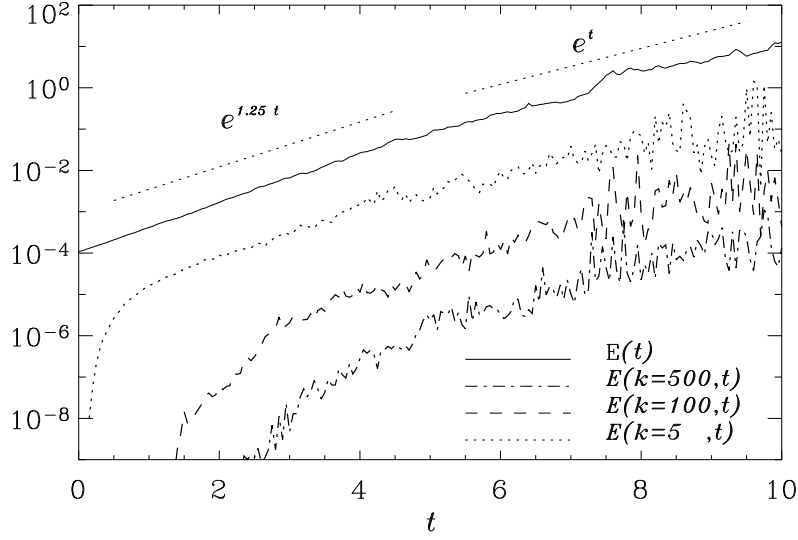}
\end{center}
\caption[]{Time growth of the total small-scale energy (solid line) and of the
energy spectrum at several fixed $k$ (dashed lines) in the case of the synthetic Gaussian strain
with a finite correlation time}
\label{fig3}
\end{figure}

\begin{figure}
\begin{center}
\includegraphics[width=.9\textwidth]{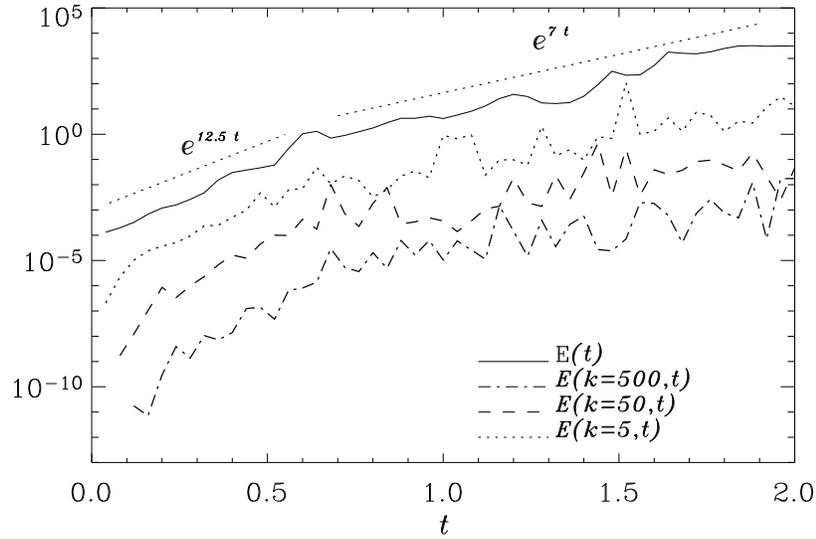}
\end{center}
\caption[]{Time growth of the total small-scale energy (solid line) and of the
energy spectrum at several fixed $k$ (dashed lines) in the case of the strain
obtained from 512$^3$ DNS}
\label{fig4}
\end{figure}

 Figures \ref{fig3} and \ref{fig4} show the time growth of the total energy and the energy
spectrum at several fixed wavenumbers 
in the first and the second numerical experiment respectively. 
One can see that the growth is approximately exponential
as predicted by the theory. It is interesting that the theory also predicts a change to a slower
exponential growth of the total energy
when time crosses $t_d$. Similar effect is called the dissipative anomaly in
the kinematic dynamo theory (\cite{kazantsev68, chertkov99}). This effect is  consistent with figure \ref{fig3} 
which shows that at $t=t_d \approx 4.5$  the slope for the total energy gets smaller and becomes
approximately equal to the slope of the individual $k$-components (as predicted by the theory).
For the second experiment the growth is also approximately exponential and the change of slope
occurs  at $t=t_d \approx 0.7$.

\begin{figure}
\begin{center}
\includegraphics[width=.9\textwidth]{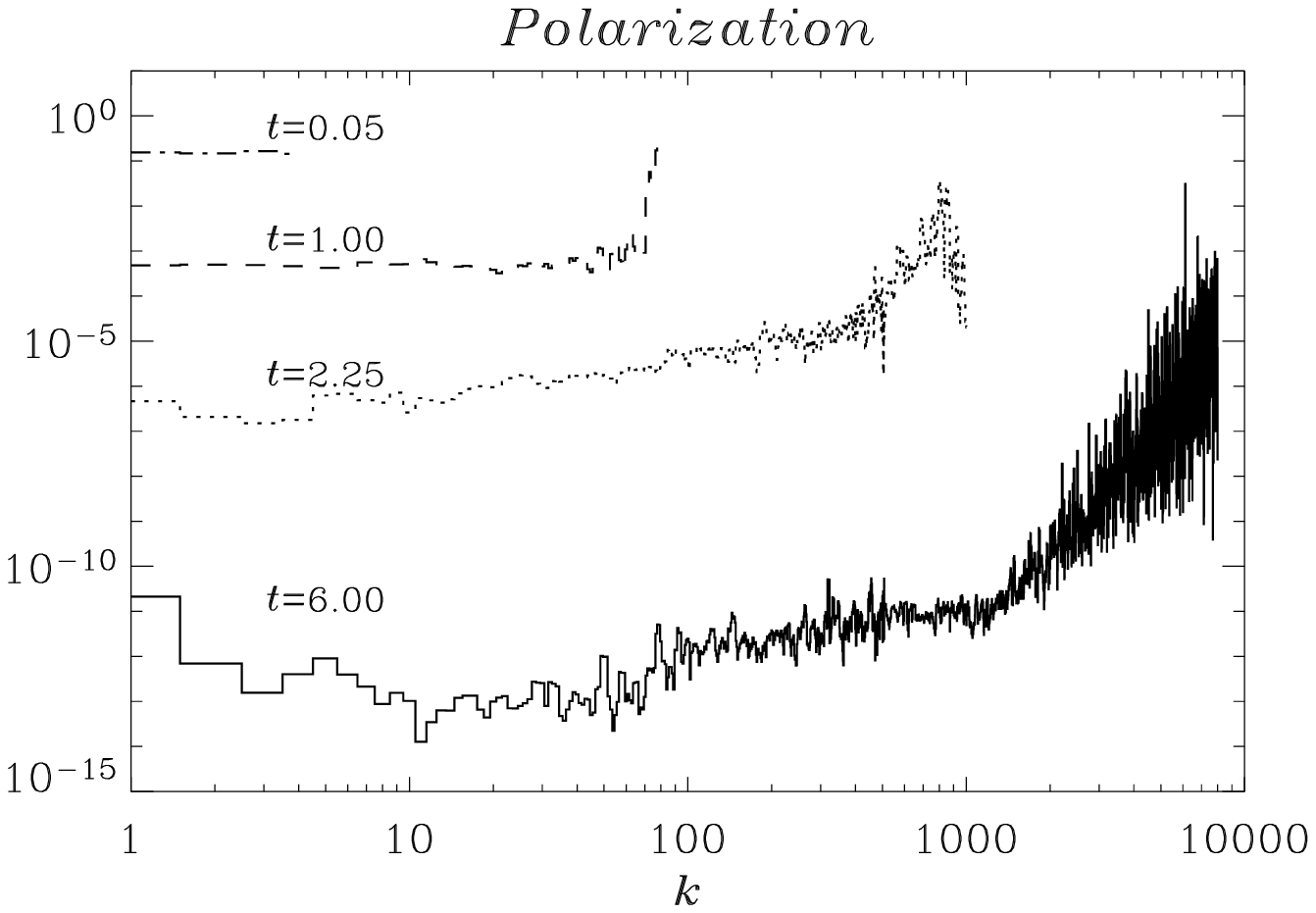}
\end{center}
\caption[]{Polarization spectrum in the case of the synthetic Gaussian strain
with a finite correlation time}
\label{fig5}
\end{figure}

\begin{figure}
\begin{center}
\includegraphics[width=.9\textwidth]{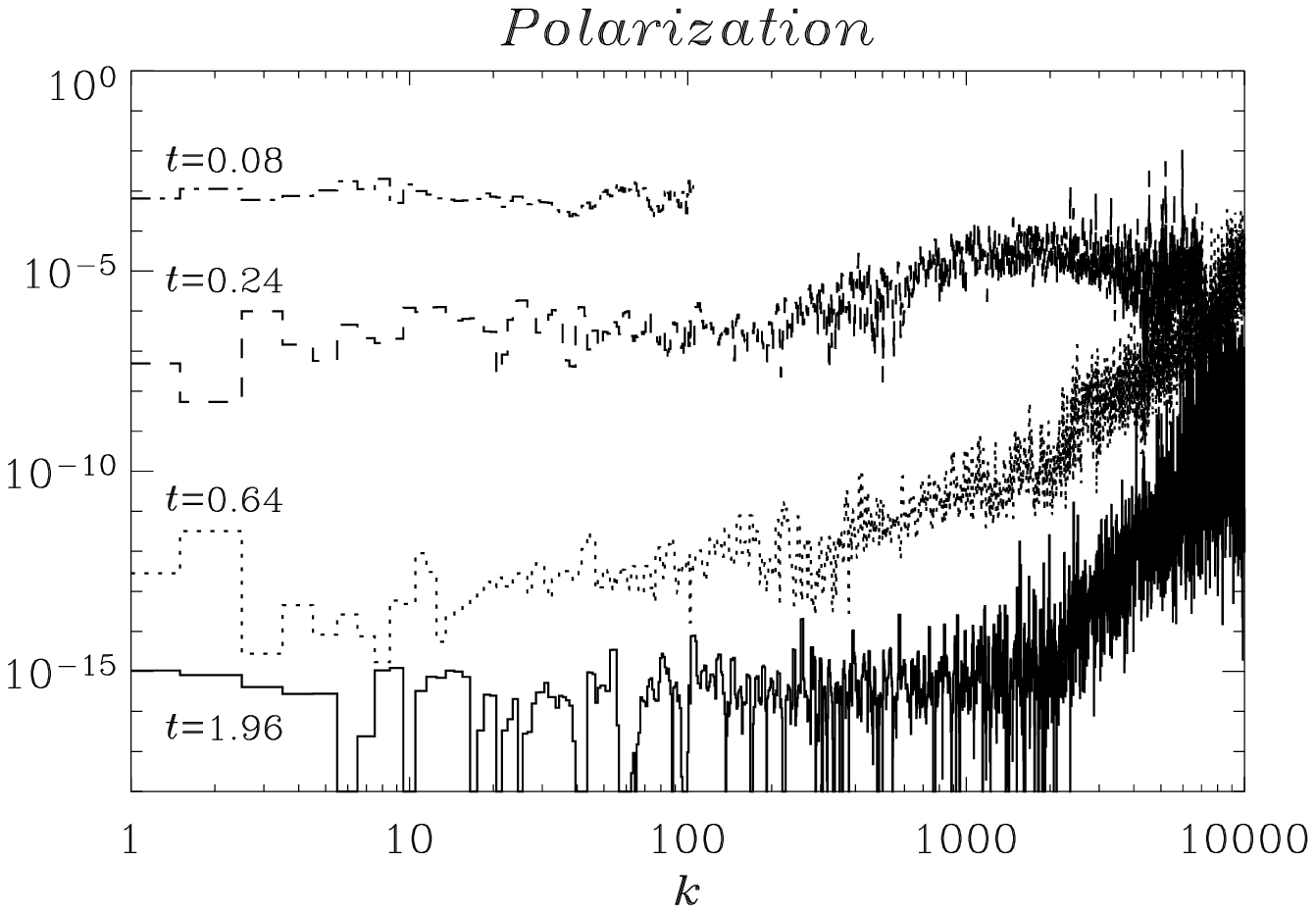}
\end{center}
\caption[]{Polarization spectrum in the case of strain
obtained from 512$^3$ DNS}
\label{fig6}
\end{figure}

Figures \ref{fig5} and \ref{fig6} show the spectrum of the polarization 
at several fixed wavenumbers 
in the first and the second numerical experiment respectively.
Similarly to the white-noise strain, in both simulations
the polarization sharply decreases in time
and it tends to an independent of $k$ spectrum in the inertial range 
(with a log-correction for $t>t_d$).

\begin{figure}
\begin{center}
\includegraphics[width=.9\textwidth]{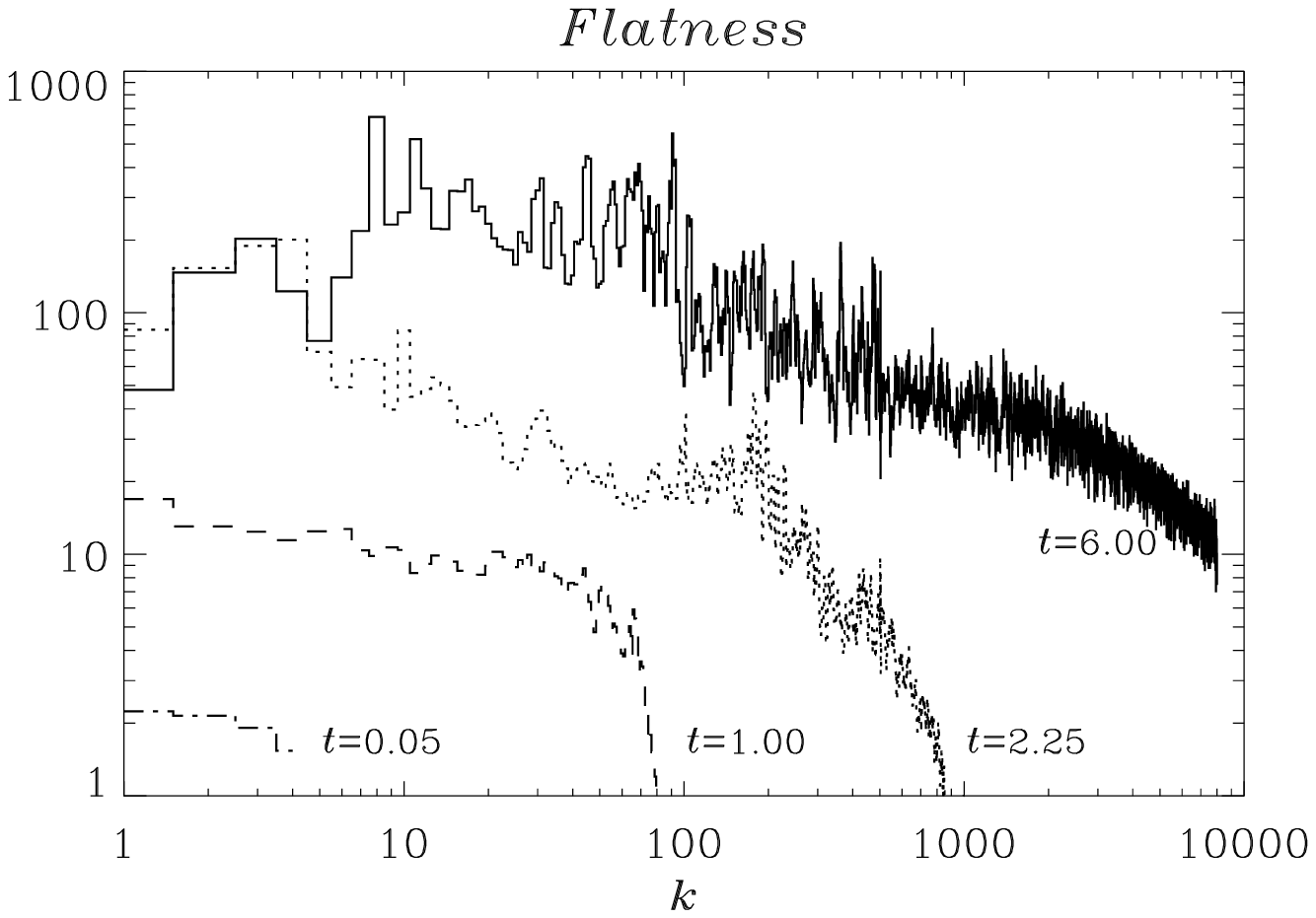}
\end{center}
\caption[]{Flatness spectrum in the case of the synthetic Gaussian strain
with a finite correlation time}
\label{fig7}
\end{figure}

\begin{figure}
\begin{center}
\includegraphics[width=.9\textwidth]{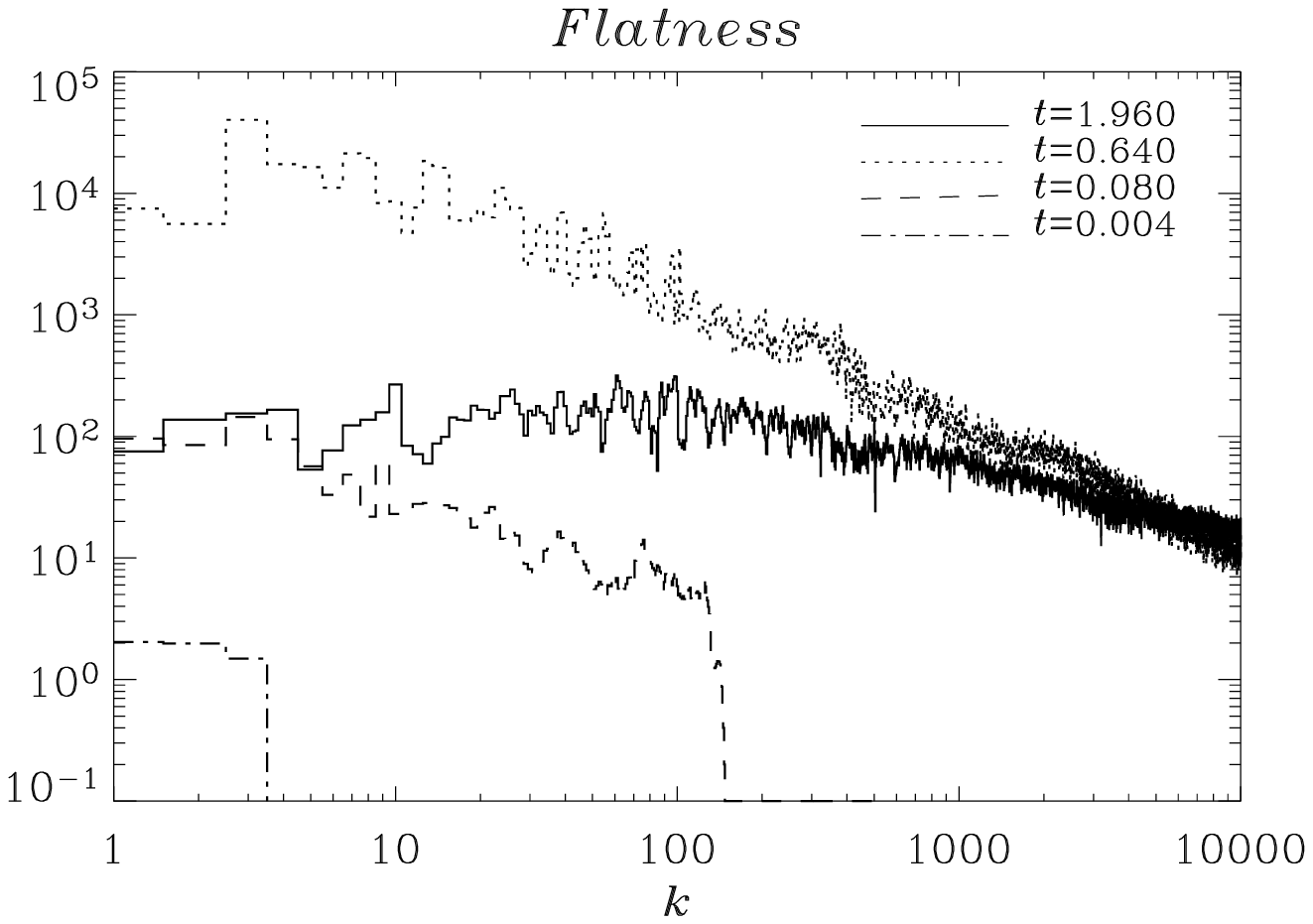}
\end{center}
\caption[]{Flatness spectrum in the case of the strain
from 512$^3$ DNS}
\label{fig8}
\end{figure}

Figures \ref{fig7} and \ref{fig8} show the spectrum of the flatness 
at several fixed wavenumbers 
in the first and the second numerical experiment respectively.
Similarly to the white-noise predictions, the flatness is growing with time.
However, the theoretical 3/2 slope is observed neither for the synthetic nor for 
the DNS strain case. The fact that these deviations are observed in the same
way for both the DNS strain and the synthetic strain (which is short correlated and
therefore quite close to the white noise process) might indicate a failure of the numerical 
method to reproduce some features of higher correlators rather than a true deviation
due to the differences in the strain statistics. Further discrepancy arises for the DNS
strain case at very large time when the flatness is observed to decrease.

\section{Conclusion}

In this paper we introduced a  description of the small-scale Navier-Stokes
turbulence with much larger scales, the SDT model. Such nonlocal interactions
dominate in 2D turbulence at large time. In 3D, they were shown to be responsible 
for intermittency in numerical experiments (\cite{laval01a}). 
The SDT model assumes that the large scales are Gaussian and short 
correlated in time and it describes turbulence in terms of
the one-point $k$-space correlators using the method introduced
in \cite{nazarenko03} for the kinematic dynamo problem.
We studied both 2D and 3D turbulence, the 2D case being equivalent to the
problem of 2D passive scalars in the Batchelor regime.
In 2D, we found steady state solutions for correlators of all orders.
These solutions correspond to forced turbulence and they describe cascades
of the energy and enstrophy series of invariants (two invariants  at each
order). The energy cascades are non-intermittent: initially Gaussian turbulence
at the forcing scale remains Gaussian and scale invariant throughout the inertial
range. On the contrary, the scale invariance and gaussianity break down for the
enstrophy cascades and a regime dominated by thin elliptical structures develops
in the $k$-space.
In 3D, we found that the steady state does not exist in SDT and the
total energy grows in a dynamo-like fashion. To have a realistic description
of the steady state of the Navier-Stokes turbulence one have to compliment SDT
with a model for the local scale interaction which will be done in future
work. However, the study of the purely
nonlocal interaction presented in this paper reveals several interesting 
effects which are likely to persist in some form when the local interactions 
are taken into account. In particular, the nonlocal interactions are shown
to lead to an interesting turbulent state in which all modes have plane
polarization in the $k$-space. Statistics of such turbulent fields is far from
being Gaussian (in which all polarizations are present).
Similar effect for the dynamo magnetic
fields was found in \cite{nazarenko03}. The $k$-space moments allow also
to quantify the dominant coherent structures in the $k$-space which
are responsible for intermittency in similar way as the  $x$-space
moments capture the $k$-space structures.
Note that singular structures in the $k$-space are not necessarily singular 
in the $x$-space (e.g. a field periodic  in some direction is a 
singular 1D line in the $k$-space).
  For the Navier-Stokes turbulence, the intermittent $k$-space
structures are found to be very elongated ellipsoids centered at the origin.
These structures leave their signatures on the spectral flatness and on the
scalings of the higher $k$-space moments with respect to the order $n$.
Further, similarly to the kinematic dynamo problem the scalings indicate
presence of the log-normal statistics of the small-scale velocity fields.
Naturally, putting back the local interactions into the model will
weaken the tendency to form elongated ellipsoids and plane polarized wavepackets,
but some deviation from the Gaussian statistics in the direction
predicted by these tendencies should be expected in real Navier-Stokes turbulence.

We studied numerically the SDT model (\ref{beqn}) for  a synthetic Gaussian strain
with a finite correlation time and for a strain field obtained from a 512$^3$ spectral
DNS of the Navier-Stokes equation. The results show that most predictions of the white-noise theory,
such as e.g. the energy spectrum shape, the polarization decrease and the increase of the spectral
flatness, are also observed in these two cases of the strain. Thus, the SDT equations
obtained for the Gaussian white noise strain are likely to be a good model for 
the nonlocal interactions in real Navier-Stokes turbulence.

\section{Appendix}

Our aim here is to derive a closed equation for the generating function $Z$
starting with equation (\ref{zdot}). The last term in this equation
is the easiest one,
\begin{equation}
 - 2 \nu k^2 \langle ({\lambda|{\bf u({\bf k})}|^2 
+ \alpha {\bf u({\bf k})}^2 + \beta {\bf \overline u({\bf k})}^2}) E
\rangle = -2 \nu k^2 {\cal D} Z
\label{last}
\end{equation}
where ${\cal D}$  is a differential operator defined in
(\ref{calD}). The correlators containing factor $ \sigma_{ij}$ can be found
 using the Gaussian integration by parts. In particular
\begin{equation}
 \langle \sigma_{ij} E \rangle  =  \Omega \langle {\delta E \over \delta \sigma_{ij} } \rangle =
 \Omega \left[ \lambda \langle (\Gamma_{m,ij} {\overline u_m} + {\overline \Gamma_{m,ij}} {u_m}) E \rangle + 
2   \alpha   \langle \Gamma_{m,ij} { u_m}  E \rangle +
2  \beta  \langle \overline \Gamma_{m,ij} {\overline u_m}  E \rangle
\right],
\label{g_int}
\end{equation}
where we have used the definition (\ref{E}). Here, $\Gamma_{m,ij}$ is a response function,
\begin{equation}
\Gamma_{m,ij} = {\delta u_m \over \delta \sigma_{ij} }.
\label{gam} 
\end{equation}
Differentiating (\ref{beqn}) with respect to $\sigma_{ij}$  and using whiteness of the
strain tensor we get
\begin{equation}
\Gamma_{m,ij} = (k_i \, \partial_j + {\delta_{ij} \over d} (1 - k_l \, \partial_l))  u_m 
+ ( {2 k_m k_i \over k^2} - \delta_{mi} )  u_j.
\label{gam1} 
\end{equation}
In what follows we will use the turbulence isotropy, in particular, expressions of type
\begin{eqnarray}
 \langle (\overline u_i u_j +  \overline u_j u_i ) E \rangle & =& 
{ 2 \over (d-1) }  \langle |{\bf u}|^2  E \rangle (\delta_{ij} - {k_i k_j \over k^2}) \\
 \langle u_i  u_j  E \rangle &=&
{ 1 \over (d-1) }  \langle {\bf u}^2  E \rangle (\delta_{ij} - {k_i k_j \over k^2}) \\
 \langle \overline u_i  \overline u_j  E \rangle &=&
{ 1 \over (d-1) }  \langle \overline {\bf u}^2  E \rangle (\delta_{ij} - {k_i k_j \over k^2}) 
\end{eqnarray}
Substituting (\ref{gam1})  into (\ref{g_int}) and using the above isotropy relations
we have
\begin{equation}
 \langle \sigma_{ij} E \rangle  =  \Omega (k_i \partial_j - {\delta_{ij} \over d}
k_l \partial_l ) Z + {2  \Omega \over (d-1)} ( {k_i k_j \over k^2} - {\delta_{ij} \over d} ) {\cal D} Z
\label{sije}
\end{equation}
where ${\cal D}$  is a differential operator defined in
(\ref{calD}).
This allows us to find the first term on the RHS of (\ref{zdot}),
\begin{equation}
k_i \partial_j \langle \sigma_{ij} E \rangle  =  \Omega \left[ {(d-1) \over d} k^2 Z_{kk} +
{1 \over d} (2 {\cal D} + d^2 -1) k Z_k + 2 {\cal D} Z  \right]
\label{1st}
\end{equation}
Similarly, the other three terms on the RHS of (\ref{zdot}) can be
obtained by the Gaussian integration by parts and using the
response function  (\ref{gam1}) and the isotropy. After a lengthy but
straightforward algebra one gets
\begin{equation}
\lambda \langle \sigma_{ml} ({\overline u_m} u_l + {\overline u_l} u_m) E
\rangle = - 2 \lambda \Omega \left[
(d - {2 \over d}) Z_\lambda + 2 (1 - {1 \over d}) {\cal D} Z_\lambda +
\lambda (Z_{\alpha \beta} - Z_{\lambda \lambda }) +
{1 \over d} k_i \partial_i Z_\lambda \right]
\label{2nd}
\end{equation}
and
\begin{equation}
 2 \alpha  \langle \sigma_{ml} { u_m} u_l E \rangle
= 2 \alpha \Omega \left[
({2 \over d} - d) Z_\alpha   + {2 \over d}  {\cal D} Z_\alpha -
{1 \over d} k_i \partial_i Z_\alpha -
2 \lambda Z_{\lambda \alpha} - 2 \beta Z_{\lambda \lambda } -
2 \alpha  Z_{\alpha \alpha}
\right]
\label{3rd}
\end{equation}
The 4th term can be obtained from (\ref{3rd}) via interchanging 
$\alpha $ with $\beta$ and ${\bf u} $ with $\overline {\bf u} $,
\begin{equation}
 2 \beta  \langle \sigma_{ml} {\overline u_m} {\overline u_l} E \rangle
= 2 \beta \Omega \left[
({2 \over d} - d) Z_\beta   + {2 \over d}  {\cal D} Z_\beta -
{1 \over d} k_i \partial_i Z_\beta -
2 \lambda Z_{\lambda \beta} - 2 \alpha Z_{\lambda \lambda } -
2 \beta  Z_{\beta \beta}
\right]
\label{4th}
\end{equation}
Putting expressions (\ref{1st}), (\ref{2nd}), (\ref{3rd}), 
(\ref{4th}) and 
(\ref{last}), we have the following final equation,
\begin{eqnarray}
\dot Z &=& \Omega \big[ (1 - { 1 \over d}) k^2 Z_{kk} + 
{ 1 \over d} (4 {\cal D} + d^2 -1) k Z_{k} 
 +2(1-{2 \over d} +d) {\cal D} Z 
- {4 \over d} {\cal D}^2 Z  \nonumber \\ &&
 + 2 (\lambda^2 +4 \alpha \beta) Z_{\lambda \lambda} +
 2 \lambda^2 Z_{\alpha \beta}  + 8 \lambda \alpha Z_{\alpha \lambda}
+ 8 \lambda \beta Z_{\beta \lambda} +4 \alpha^2 Z_{\alpha \alpha} 
+4 \beta^2 Z_{\beta \beta}
\big] \nonumber \\ &&
-2 \nu k^2 {\cal D} Z,
\label{zdotFV}
\end{eqnarray}


\begin{thebibliography}{18}
\expandafter\ifx\csname natexlab\endcsname\relax\def\natexlab#1{#1}\fi

\bibitem[Balkovsky \& Fouxon(1999)]{balkovsky99}
{\sc Balkovsky, E. \& Fouxon, A.} 1999 Universal long-time properties of
  lagrangian statistics in the batchelor regime and their application to the
  passive scalar problem. {\em Phys. Rev. E\/} {\bf 60}, 4164--4174.

\bibitem[Batchelor \& Proudman(1954)]{batchelor54}
{\sc Batchelor, G.~K. \& Proudman, I.} 1954 The effect of rapid distortion of
  fluid in turbulent motion. {\em Quart. Journ. Mech. and Applied Math.\/} {\bf
  7}, 83--103.

\bibitem[Chertkov {\em et~al.\/}(1999)Chertkov, Falkovich, Kolokolov \&
  Vergassola]{chertkov99}
{\sc Chertkov, M., Falkovich, G., Kolokolov, I. \& Vergassola, M.} 1999 Small
  scale turbulent dynamo. {\em Phys. Rev. Lett.\/} {\bf 83}, 4065--4068.

\bibitem[Falkovich {\em et~al.\/}(2001)Falkovich, Gawedzki \&
  Vergasolla]{falkovich01}
{\sc Falkovich, G., Gawedzki, K. \& Vergasolla, M.} 2001 Particles and fields
  in fluid turbulence. {\em Rev. Mod. Phys.\/} {\bf 73}, 913--975.

\bibitem[Furstengerg(1963)]{furstenberg63}
{\sc Furstengerg, H.} 1963 Non commuting random products. {\em Trans. Am. Math.
  Soc.\/} {\bf 108}, 377--428.

\bibitem[Kazentsev(1968)]{kazantsev68}
{\sc Kazentsev, A.~P.} 1968 Enhancement of a magnetic field by a conducting
  fluid. {\em Sov. Phys. JETP\/} {\bf 26}.

\bibitem[Kraichnan(1961)]{kraichnan61}
{\sc Kraichnan, R.~H.} 1961 Dynamics of nonlinear stochastic systems. {\em J.
  Math. Phys.\/} {\bf 2}, 124--148.

\bibitem[Kraichnan(1974)]{kraichnan74}
{\sc Kraichnan, R.~H.} 1974 Convection of a passive scalar by a quasi-uniform
  random straining field. {\em J. Fluid Mech.\/} {\bf 64}, 737--762.

\bibitem[Kraichnan \& Nagarajan(1967)]{kraichnan67b}
{\sc Kraichnan, R.~H. \& Nagarajan, S.} 1967 Growth of turbulent magnetic
  fields. {\em Phys. Fluids\/} {\bf 10}, 859--870.

\bibitem[Kulsrud \& Anderson(1992)]{kulsrud92}
{\sc Kulsrud, R. \& Anderson, S.~W.} 1992 The spectrum of random magnetic
  fields in the mean field dynamo theory of the galactic magnetic field. {\em
  Astrophysical Journal\/} {\bf 396}, 606.

\bibitem[Laval {\em et~al.\/}(2001)Laval, Dubrulle \& Nazarenko]{laval01a}
{\sc Laval, J.-P., Dubrulle, B. \& Nazarenko, S.} 2001 Non-locality and
  intermittency in {3D} turbulence. {\em Phys Fluids\/} {\bf 13}, 1995--2012.

\bibitem[Nazarenko(1999)]{nazarenko00c}
{\sc Nazarenko, S.} 1999 Exact solutions for near-wall turbulence theory. {\em
  Phys. Rev. A\/} {\bf 264}, 444--448.

\bibitem[Nazarenko {\em et~al.\/}(2000)Nazarenko, Kevlahan \&
  Dubrulle]{nazarenko00d}
{\sc Nazarenko, S., Kevlahan, N. K.-R. \& Dubrulle, B.} 2000 Nonlinear {RDT}
  theory of near-wall turbulence. {\em Physica D\/} {\bf 139}, 158--176.

\bibitem[Nazarenko \& Laval(2000)]{nazarenko00b}
{\sc Nazarenko, S. \& Laval, J.-P.} 2000 Non-local 2{D} turbulence and
  {B}atchelor's regime for passive scalars. {\em J. Fluid Mech.\/} {\bf 408},
  301--321.

\bibitem[Nazarenko {\em et~al.\/}(2003)Nazarenko, West \&
  Zaboronski]{nazarenko03}
{\sc Nazarenko, S., West, R.~J. \& Zaboronski, O.} 2003 Statistics of fourier
  modes in the kazantsev-kraichnan dynamo model. {\em Submitted to Phys. Rev.
  E\/} .

\bibitem[Orszag(1966)]{orszag66}
{\sc Orszag, S.~A.} 1966 Dynamics of fluid turbulence. {\em Tech. Rep.\/}. PPPL
  report PPL-AF-13.

\bibitem[Schekochihin {\em et~al.\/}(2002{\natexlab{{\em a\/}}})Schekochihin,
  Boldyrev \& Kulsrud]{schekochihin02a}
{\sc Schekochihin, A.~A., Boldyrev, S.~A. \& Kulsrud, R.~M.}
  2002{\natexlab{{\em a\/}}} Spectra and growth rates of fluctuating magnetic
  fields in the kinematic dynamo theory with large magnetic prandtl numbers.
  {\em Astrophysical Journal\/} {\bf 567}, 828, e-print astro-ph/0103333.

\bibitem[Schekochihin {\em et~al.\/}(2002{\natexlab{{\em b\/}}})Schekochihin,
  Maron, Cowley \& McWilliams]{schekochihin02b}
{\sc Schekochihin, A.~A., Maron, J.~L., Cowley, S. \& McWilliams, J.~C.}
  2002{\natexlab{{\em b\/}}} The small-scale structure of magnetohydrodynamic
  turbulence with large magnetic prandtl numbers. {\em Astrophysical Journal\/}
  {\bf 576}, 806--813, e-print astro-ph/0203219.

\end{thebibliography}

\end{document}